\documentclass{article}
\usepackage{icrctc07}

\title{Performance of  the Pierre Auger Observatory Surface Detector}
\shorttitle{Performance of the Pierre Auger Observatory Surface Detector}

\authors{T.Suomij\"{a}rvi$^1$, for the Pierre Auger Collaboration$^2$} 

\shortauthors{Pierre Auger Collaboration}
\afiliations{(1) Institut de Physique Nucleaire, Universit\'{e} Paris-Sud, IN2P3-CNRS,  Orsay, 
 France \\
(2) Observatorio Pierre Auger, Av. San Mart\'{i}n Norte 304, (5613)
Malarg\"{u}e, Argentina} 
\email{tiina@ipno.in2p3.fr}

\abstract{The Surface Detector of the Pierre Auger Observatory will consist of 1600 water Cherenkov tanks sampling ground particles of air showers produced  by energetic cosmic rays. The arrival times are obtained from GPS and power is provided by solar panels. The construction of the array is nearly completed and a large number of detectors has been operational for more than three years.
In this paper the performance of different components of the detectors are discussed. The accuracy of the signal measurement and the trigger stability are presented. The performance of the solar power system and other hardware, as well as the water purity and its long-term stability are discussed. }

\begin{document}
\maketitle
\section{Introduction}

The Surface Detector (SD) of the Pierre Auger Observatory is composed of
1600 water Cherenkov detectors extending over an area of 3000~km$^2$ with 1500~m spacing between detectors. 
The detectors are cylindrical 
polyethylene tanks with 3.6~m diameter and about 1.6~m height.  The 
inside of the tank is covered by a Tyvek$^{\textregistered}$ liner 
for uniform reflection of the Cherenkov light, and the liner is 
filled with purified water of quality typically better than  5~M$\Omega$.cm, up 
to a depth of 1.2~m.   Two solar panels combined with 
batteries furnish 10~W power for the station.

The signals produced by the Cherenkov light are read out by three large 9'' XP1805 Photonis 
photomultipliers (PMTs).  The PMTs are equipped with a resistive base
having two outputs: anode and amplified last dynode. This allows a 
large dynamic range, totaling  15 bits, extending from a few to about $10^5$ 
photoelectrons. The high voltage is provided locally. The nominal operating 
gain of the PMTs is 2$\times$10$^5$ and can be extended to 10$^6$.  The base, together with the HV 
module, is protected against humidity by silicone potting. 

The signals from anode and dynode are filtered  and digitised at
40~MHz using 10 bit Flash Analog-Digital Converters (FADC). Two
shower triggers are used: threshold trigger (ThT) and time-over-threshold (ToT)
trigger. The first one is a simple majority trigger with a threshold 
at 3.2 VEM (Vertical Equivalent Muon) in each of 4 or more nearby detectors. The 
ToT trigger requires 12 FADC bins with signals larger 
than 0.2 VEM in a sliding window of 3~$\mu$s in each of 3 or more
detectors in a compact configuration. The time-over-threshold trigger efficiently triggers on the 
shower particles far away from the shower core. In addition, a muon 
trigger allows for recording of  continuous calibration data.  

 A common time base is established for different detector
stations by using the GPS system. Each tank is equipped with a 
commercial GPS receiver (Motorola OnCore UT) providing a one pulse 
per second output and software corrections. This signal is used to 
synchronise a 100~MHz clock which serves to timetag the trigger. Each 
detector station has an IBM 403 PowerPC micro-controller for local
data acquisition, software trigger and detector monitoring, and memory 
for data storage. The station electronics is 
implemented in a single module called the Unified Board, and mounted in 
an aluminum enclosure. The electronics package is mounted on top of the hatch 
cover of one of the PMTs and protected against rain and dust by an 
aluminum dome. 

The detector calibration is inferred from background muons. The
typical rise time for a muon signal is about 15~ns with a decay time of 
the order of 70~ns. The average number of photoelectrons per muon 
collected by one PMT is 95. By adjusting the trigger rates, the PMT 
gains are matched within 6\%. The measurement of the muon charge
spectrum allows us to deduce the charge value for the Vertical Equivalent Muon, $Q_{VEM}$, 
from which the calibration is inferred for the whole dynamic range. The cross 
calibration between the two channels, anode and dynode outputs, is performed by 
using small shower signals in the overlap region of the two channels \cite{calibnim}. 

The decay constant of the muon signal is related to the absorption length of the light produced. This depends of various parameters such as the Tyvek$^{\textregistered}$ reflectivity and the purity of the water. The signal decay constant correlates with the so called area-to-peak (A/P) ratio of the signal: 
\begin{equation}
{\rm A/P} = \frac{Q_{VEM}}{I_{VEM}}
\end{equation}

where $I_{VEM}$ is the maximum current of the muon signal. This area-to-peak ratio is a routine monitoring quantity that is directly available from the local station software. More complete description of the Pierre Auger Observatory surface detector can be found in ref. \cite{EANIM}.

\section{Uniformity of the detector responses}

The stable data taking with the Surface Detector started in January
2004 and various parameters are continuously monitored to ensure the 
good performance of the detectors. The noise levels are very low. 
For both the anode and dynode channels, 
the mean value of the pedestal fluctuation RMS is below 0.5 FADC channels corresponding to about 0.01 VEM.
The intrinsic resolution of the GPS time tagging system is about 8~ns 
requiring a good precision for the station location. An accuracy better than
1 meter is obtained for the tank position  by measuring the positions with differential GPS. 

\begin{figure}
\begin{center}
\noindent
\includegraphics [width=0.5\textwidth]{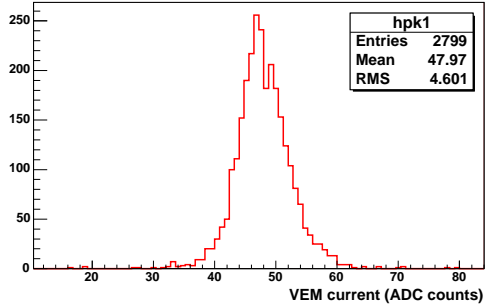}
\end{center}
\caption{VEM measured for 2799 PMTs.}\label{fig:VEM}
\end{figure}

\begin{figure}
\begin{center}
\noindent
\includegraphics [width=0.5\textwidth]{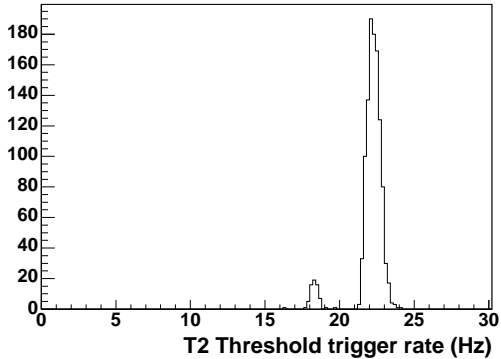}
\end{center}
\caption{Threshold  trigger rates for the SD stations.}\label{fig:trigger_th}
\end{figure}

\begin{figure}
\begin{center}
\noindent
\includegraphics [width=0.5\textwidth]{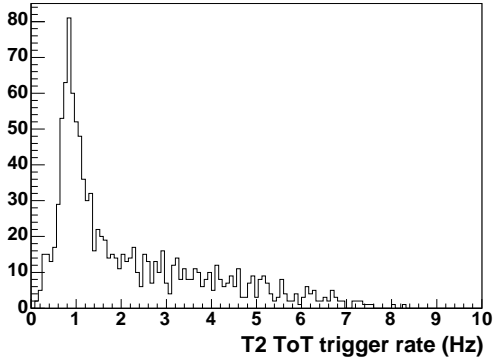}
\end{center}
\caption{Time-over-threshold trigger rates for the SD stations.}\label{fig:trigger_tot}
\end{figure}

 Figure \ref{fig:VEM} shows the muon peak ($I_{VEM}$)  values for the SD. The mean value is at channel 48 with an RMS of 4.6 showing a very good uniformity in the detector response.  Trigger rates (Figs. \ref{fig:trigger_th}, \ref{fig:trigger_tot} ) are also remarkably uniform over all detector 
stations, also implying good calibration and baseline determination. The mean value of the threshold trigger is 22 Hz with dispersion less than 2\%. The time-over threshold trigger is about 1 Hz with a larger dispersion. This is due to the fact that it depends on the charge  and is more sensitive to the characteristics of the tank. It is observed that the new tanks often have ToT values which are higher and then stabilise after a few months to about 1 Hz.  The trigger studies and the studies on the muon response show that the detectors have, after a few month stabilisation, a very good uniformity. 

The day-night atmospheric temperature  variations can be larger than 20 $^\circ$C.  In each tank, temperature is measured on the PMT bases, on the electronics board, and on the batteries which allows to  correlate various monitoring data with the temperature.  The typical day-night variations are of the order of 2 ADC channels for the muon peak. This is mainly due to the sensitivity of the PMTs to temperature. These temperature variations 
also slightly affect the ToT-trigger. The muon calibration is made  on-line every minute. This continuous calibration allows to correct for the day-night temperature effects.  

\section{Long-term performance}

\begin{figure}
\begin{center}
\noindent
\includegraphics [width=0.52\textwidth]{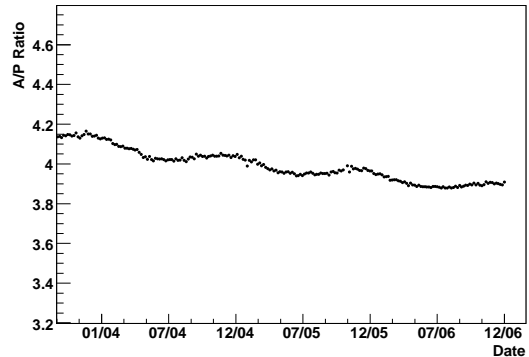}
\end{center}
\caption{Area-to-peak ratio as a function of time for a typical PMT channel.}\label{fig:AreaPeak_time_Option1}
\end{figure}

Figure \ref{fig:AreaPeak_time_Option1} shows the area-to-peak ratio for a typical PMT channel as a function of time. Two effects are observed: a slight global decrease and small seasonal variations. The decrease of the signal as a function of time could be due to changes in the liner reflectivity or in the water quality.

The time dependence of the area-to-peak ratio is fitted with a function that
contains the observed bulk properties of stations in this early phase
of operation.  A goodness-of-fit test is used to decide between two possible
functions.  In the first, a decay term is included as an estimate of the likely
change in the ratio over the long term while the second assumes a pure linear
decrease over time.
\begin{eqnarray*}
\mbox{\rm A/P} &=& p_{0}\Big[1-p_{1}\cdot (1-e^{\frac{-t}{p_{2}}})\Big]\\
      & & \times\Big[1+p_{3}\cdot \sin(2\pi(\frac{t}{T}-\phi))\Big] \\
\mbox{\rm A/P} &=& p_{0}\Big[1-p_{1}\cdot t\Big]\\
      & & \times\Big[1+p_{3}\cdot \sin(2\pi(\frac{t}{T}-\phi))\Big]
\end{eqnarray*}

The fit parameters are $p_{0}$ through $p_{3}$ with the following
definitions:  $p_{0}$ is the overall normalization and  $p_{1}$ represents the fractional signal loss.
$p_{2}$ is the characteristic time and $p_{3}$ is the seasonal amplitude.  The  variable $t$ is the time in years. 

\begin{figure}
\begin{center}
\noindent
\includegraphics [width=0.52\textwidth]{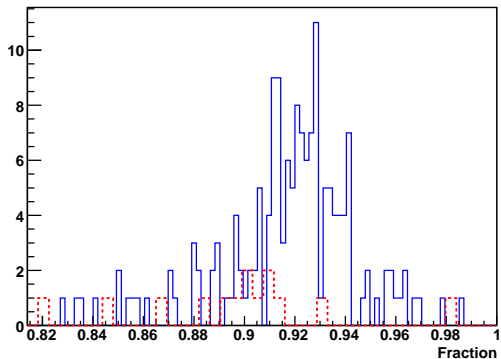}
\end{center}
\caption{Remaining signal after 10 years of operation.  The solid line is for stations that are exponential in time while the dashed lines are for stations linear in time.}\label{fig:FractionalLoss}
\end{figure}

75 stations have been in continuous operation since September 2003, yielding
samples from 225 individual PMTs.  80\% pass
quality cuts during the fitting procedure with 18 best represented
by the more worrisome linear decrease with time.  Those same 18 are among
the lowest in terms of fractional loss, see figure \ref{fig:FractionalLoss}, so that the
net result after 10 years is similar regardless of whether a stable
point is predicted or not.

The stabilization timescales predicted from these fits to a simple model
range from a fraction of a year out to 25 years, with the majority achieving
stability during the first several years of operation. Seasonal amplitudes
are $\sim$1\% of the nominal area-to-peak ratio with no correlation to the
characteristic time. The expected fractional signal loss in 10 years is less than 10\% which gives confidence in a very stable long term operation.

\section{Maintenance}

As of May 2007, more than 1200 surface detector stations are
operational. Typically more than 98.5\% of the stations are
operational at any time.
Only seven liners were observed to leak shortly after installation.
In these cases, which constitute the worst failure mode, the tank is
emptied and brought back to the Assembly Building for replacement of
the interior components. During more than three years of operation,
only 12 solar panels have been damaged or were missing.
Solar power system parameters are recorded and analyzed using the
central data acquisition system.
The number of electronic failures is small. Only about 20 electronic kits have been brought back for repair.  Some failures have been detected due to bad connections. The 
PMT failures are less than 1\%. The maintenance rate is within original expectations.

\section{Conclusions}

In conclusion, with over 1200 detectors in the field, some of which
have already been operational for over five years, much insight on
their performance has been gained. The design is robust to withstand
the adverse field conditions and failure rates are less than
expected. Data taking is ongoing and the first scientific results
have already been published. The physics
performance has met or exceeded all of our requirements.

\bibliography{icrc0299}
\bibliographystyle{plain}
\end{document}